\documentclass[journal]{IEEEtran}

\usepackage{amsmath,amsfonts}
\usepackage{algorithmic}
\usepackage{algorithm}
\usepackage{array}
\usepackage{textcomp}
\usepackage{stfloats}
\usepackage{url}
\usepackage{verbatim}
\usepackage{graphicx}
\usepackage{cite}
\usepackage{color}
 \usepackage{bm}
\usepackage{float}  
\usepackage{multirow}
\usepackage{diagbox} 
\usepackage{multirow} 
\usepackage{threeparttable}
\usepackage{stfloats}
\usepackage{subfigure}
\usepackage{soul}

\begin{document}

\title{Digital Twin Channel for 6G: Concepts, Architectures and Potential Applications}

\author{Heng Wang, Jianhua Zhang, Gaofeng Nie, Li Yu, Zhiqiang Yuan, Tongjie Li, Jialin Wang, Guangyi Liu
}

\markboth{Journal of \LaTeX\ Class Files,~Vol.~14, No.~8, August~2021}%
{Shell \MakeLowercase{\textit{et al.}}: A Sample Article Using IEEEtran.cls for IEEE Journals}


\maketitle

\begin{abstract}

Digital twin channel (DTC) is the real-time mapping of radio channels and associated communication operations from the physical world to the digital world, which is expected to provide significant performance enhancements for the sixth-generation (6G) communication system. This article aims to bridge the gap between conventional channel twin research and emerging  DTC, by defining five evolution levels of channel twins from aspects including methodology, data category, and application. Up to now, the industry and academia have made significant progress in the fourth-level twin, and have begun the research on the fifth-level twin, i.e., autonomous DTC, which offers the opportunity for a new 6G communication paradigm. This article subsequently provides detailed insights into the requirements and possible architecture of a complete DTC for 6G. Then, the feasibility of real-time DTC is experimentally validated. Finally, drawing from the 6G typical usages, we explore the potential applications and the open issues in future DTC research.

\end{abstract}



\section{Introduction}
\IEEEPARstart{T}{he} sixth-generation (6G) communication system is expected to offer intelligent, hyper-reliable, and ubiquitous connectivity, supporting the Internet of Everything (IoE) trend for future digital society. Heading towards this envisaged outlook, \textit{self-sustaining} and \textit{proactive-online learning-enabled} wireless systems are required \cite{khan2022digital}. The digital twin technique is considered an innovative tool to enable these features for 6G. By creating a real-time replica of the physical entities in the digital world, the digital twin network (DTN) can offer up-to-date network status, closed-loop decisions, and real-time interaction between the digital world and physical world, achieving proactive adaption to the complex wireless network for 6G \cite{khan2022digital}\cite{lin20236g}.

To realize the vision of DTN, accurate radio channel characteristics between the transmitter (Tx) and receiver (Rx) are required, encompassing propagation parameters, fading conditions, and other relevant properties.  {In this article, we define the mapping of radio propagation behaviors and associated physical layer operations as channel twin. It can facilitate the decision-making for various wireless tasks such as system optimization, resource allocation, and communication operations, forming the basis of DTN. \textcolor{black}{Channel models are the conventional methods to implement channel twin, albeit without decision-making.} They are the mathematical expressions characterizing the radio channel, \textcolor{black}{based on which the propagation characteristics of interest can be reproduced.} However, the conventional channel modeling methods are generally incompetent in creating the channel twin required by DTN. For instance, the widely employed stochastic modeling method is based on the statistical analysis of extensive channel measurements, and the resulting channel twin cannot offer real-time channel status and decision-making for a specific moment and site\cite{zeng2021toward}. To make DTN a reality, high-fidelity and efficient channel characterization is necessary, leading to the emergence of a real-time digital twin channel (DTC). As defined in \cite{wang2023towards}, DTC is a novel implementation of the channel twin that comprehensively reflects the realistic physical channel characteristics within the digital world, thereby enabling proactive decision-making for physical communication entities. Moreover, by leveraging the feedback from physical entities, DTC keeps learning and evolving to improve the accuracy of digital mapping, making communication operations increasingly intelligent and reliable. The substantial capabilities of DTC in channel characterization and decision-making form the foundation for a DTN-based 6G network.

\par Up to now, some research on DTC has been published\cite{wang2023towards,liang2024data,zhang2023ai,alkhateeb2023real,nie2022predictive}. However, the relationship between conventional channel research and the emerging DTC remains unresolved. Motivated by this problem, this article attempts to bridge the gap between conventional channel research and DTC by establishing a framework of five evolution levels of channel twins. Furthermore, the concept, definition, and application visions of DTC are discussed from independent and different perspectives. For instance, \cite{wang2023towards} emphasizes the explorations of radio environment knowledge in DTC. For DTC implementation, \cite{liang2024data} proposes a high-accuracy channel prediction method based on the convolutional time-series generative adversarial network. The applications of DTC for predictive 6G networks are elaborated in \cite{nie2022predictive}. This article aims to present a comprehensive roadmap, incorporating the requirements, architecture, and key modules for constructing a complete DTC. Finally, in alignment with the 6G vision, this article explores the potential DTC applications across typical usage scenarios.

 \section{Evolution of Channel Twin}

This section attempts to integrate the conventional channel research and emerging DTC into a unified framework. For this target, we define five levels of channel twins from multi-dimensional perspectives, as summarized in Table \ref{comparison}.

\begin{table*}[!t]
\centering
    \caption{The Five Levels of Channel Twins}
\label{comparison}
\begin{tabular}{|c|c|c|c|c|c|}
\hline
Attributes           & $L_1$: \!Analytic twin
  & $L_2$:\! Primary twin 
 & $L_3$:\! Intermediate twin & $L_4$:\! Advanced twin & $L_5$:\! Autonomous twin \\
 \hline
\!Physical environment\!
           & Exclude & Partially include & Include& Include & Include  \\ \hline
Sensing     &   Manual   &  Manual                    &  Manual &  Automatic&  Automatic  \\ \hline
Application
 &  Offline
   &  Offline                  & Offline   &  Online                  & Online \\ \hline
Methodology
      & \!Mathematical derivation\!
&\!Data-based statistics\! 
& Data-based AI\! / \!RT& DMDD \& AI & DMDD \& \!enhanced AI 
 \\ \hline
Data category
      & N/A & Channel&Channel\! \&\! map\!&\!Channel \!\&\! sensing \!\& map\!&  All data
      \\ \hline
 Data source
      & N/A
 & Single-node & Single-node& Single-node& Multi-node
 \\ \hline
 
Real-time decision
      & No  & No & No & Yes&Yes
      \\   \hline 
 Model update
      & Manual  & Manual & Manual& Manual&Automatic
      \\   \hline 
      
\end{tabular}
\label{table}
\end{table*}

 \subsection{Analytic Twin}

Level-$1$ ($L_1$) channel twin is generated from the mathematical formulation with specific assumptions, while actual channel measurements are absent. Thus, it is defined as an analytic twin. Representative examples of such twins include the popular Rice and Rayleigh channel generations, which have laid a solid foundation for wireless communication. Though widely employed in communication capacity analysis and system optimization, it leaves out the environment features and is incompetent for assisting in site-specific communication decision-making.

 \subsection{Primary Twin}
The first sliding correlation-based channel sounder proposed in $1970$s offers an opportunity for actual channel measurements. Based on the statistical analysis of massive channel data, one can formulate a distribution model that describes the statistical channel characteristics in typical scenarios. The standardized 5G 3GPP 38.901 and ITU-R M.2412 models serve as the representative. It relies on statistical analysis in macro-level environments, such as urban micro (UMi) and urban macro (UMa) environments, rather than considering a specific propagation environment. In other words, it only incorporates partial environmental information. As a result, the obtained channel twin may deviate significantly from the actual channel for site-specific scenarios, thereby offering limited decision-making assistance for real-time communication operations. This channel characterization is defined as the primary twin, i.e., level-2 ($L_2$) twin.

 \subsection{Intermediate Twin}
At this level, environment features are started to be taken into account for site-specific channel characterization. The ray-tracing (RT) method is capable of accurately computing channel characteristics with detailed three-dimensional (3D) environment maps. On the other hand, combined with an environment map, one can mine channel evolution patterns and subsequently predict the realistic channel characteristics via artificial intelligence (AI) techniques \cite{nie2022predictive}. Note that the channel prediction discussed in this article extends beyond predicting channels at subsequent sites; it also involves inferring the channel conditions at the current site. Both RT and AI methods are competent in reproducing the site-specific channel characteristics, which offers the possibilities for reliable decision-making. However, since they are executed offline, they cannot capture the dynamically changing environment and channel characteristics, thus failing to provide real-time assistance. The resulting channel twin is incomplete for DTN. Thus, they are defined as level-$3$ ($L_3$) twin, i.e., intermediate twin.

\par 

 \subsection{Advanced Twin}
\par 

With the multi-modal sensing and data fusion techniques being in the ascendant, more data can be collected and utilized to improve the AI-based channel prediction. The environment information extracted from sensing data enhances the generalization of channel twins. Furthermore, as discussed in \cite{wu2020artificial}, low-dimensional environment features also can contribute to improving channel prediction accuracy without requiring extensive complex computations. Based on extensive measurements, various knowledge (propagation law) of radio propagation has been abstracted in $L_2$ twin, such as the spatial consistency model and blockage probability model, among others. They are indeed the physical constraints, which can not only assist in AI network training but also significantly improve the interpretability of twins. Thus, one can establish a data-model dual-driven (DMDD) AI-based framework to realize a more efficient channel twin. 

Specifically, the wireless devices collect the multi-modal channel and environment data from distributed infrastructures, which are then fed into the digital world. After data processing, one can employ the AI module to predict the channel characteristics of interest. Subsequently, the obtained channel characteristics are utilized to make decisions for wireless operations (e.g., beam selection) in the next moment (beyond the channel coherent time), aiming to assist device communications. This twin is defined as the level-$4$ ($L_4$) twin, i.e., advanced twin. The essence is that all the aforementioned procedures are executed in real time and online. Some publications have presented the advancements on $L_4$ twin. For example, \cite{yang2023environment} introduces a beam prediction task that leverages environment information as input for training the AI network. By exploiting the channel semantics hidden in environment images, this approach offers reliable decision-making assistance with significantly reduced system overheads.

\begin{figure*}[!t]
	\setlength{\abovecaptionskip}{-0.05cm}   

\centering
\includegraphics[width=6.7in]{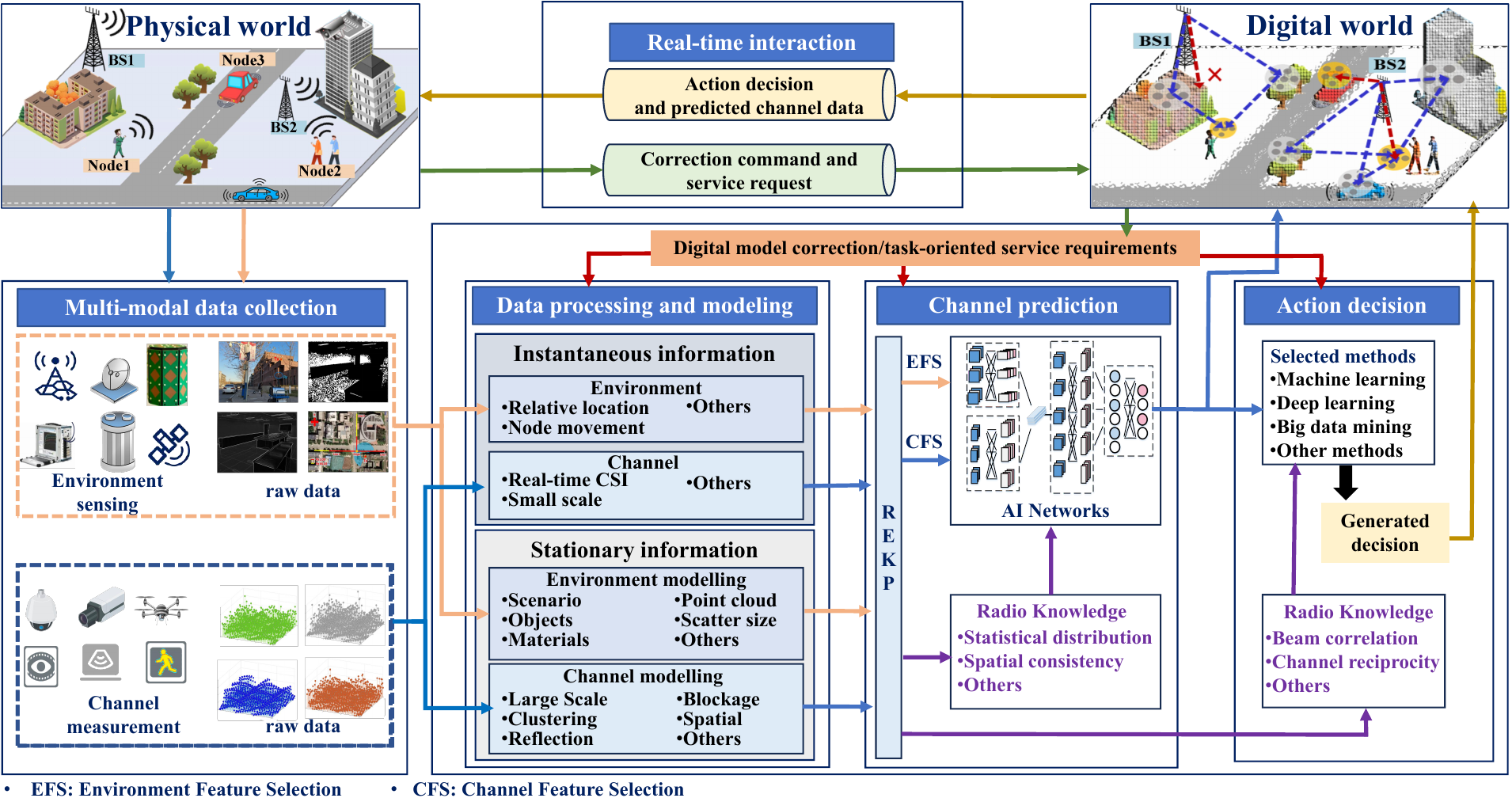}
\caption{The structure diagram and key modules of DTC.}
\label{fig_1}
\end{figure*}
 \subsection{Autonomous Twin}
The DTC discussed in this article can be defined as the level-$5$ ($L_5$) twin, with which the digital world and the physical world interact in real time. The core advantages of this technique lie in its “real-time” feature, which is lacking in the $L_1$ to $L_3$ twins, and its “self-learning” capability, which is absent in the $L_1$ to $L_4$ twins. On the one hand, the digital world provides decision-making assistance for physical devices to improve communication performance. On the other hand, according to the feedback from the physical world and prior decisions, DTC keeps self-sustaining learning and updating to improve the accuracy of digital mapping, thereby enhancing its reliability. This groundbreaking channel twin is referred to as an autonomous twin. It enables self-sustaining and proactive-online learning wireless systems and unlocks the potential for novel capabilities and applications.

\par  The key enablers of autonomous twin encompass high-fidelity sensing and enhanced AI. Specifically, \textcolor{black}{enhanced AI refers to an AI network that is capable of} integrating vast amounts of data from numerous wireless nodes and exhibiting self-learning capabilities. Ultimately, DTC is expected to evolve into a more centralized and cooperative structure\cite{alkhateeb2023real}, where multiple wireless nodes share sensing and channel information. This centralized DTC can jointly optimize decision-making among multiple nodes, thereby benefiting dense network communication. Moreover, in future autonomous DTC, not only the channel and sensing information can be utilized, but also all the available information, such as historical data, network status information, and user preferences, will be exploited to aid DTC in its learning and evolution.

 \section{How to Realize DTC}

In this section, we elaborate on the requirements, architecture, and modules of DTC. Next, we experimentally validate the feasibility of real-time environment sensing, reconstruction, and channel prediction with our designed architecture.

 \subsection{Key Requirements}
To fully leverage the role of DTC in the 6G network, the following requirements should be satisfied:

 \textbf{Precise mapping:} The foundation of DTC lies in a realistic and comprehensive replica of the physical world, comprising not only the 3D environment attributes but also the channel characteristics and transceiver operations. It should precisely capture and reflect the core features of physical objects for reliable decision-making.

 \textbf{Real-time:} Synchronization between the physical world and the digital world, is of great significance for DTC. This necessitates time-efficient executions for data sensing and processing, channel prediction, and decision-making. Therefore, the DTC architecture should be carefully designed, striking a balance of performance and complexity.

 \textbf{Self-update:} The constructed DTC may not perfectly replicate the physical entities. Various flaws may exist within the entire framework, including sensor errors, information losses, and model defects, possibly resulting in non-optimal or even incorrect decisions for physical communication devices. This demands the ability to self-update for DTC.

 \textbf{Task-oriented:} Since different communication tasks typically rely on different channel characteristics, it leads to distinct requirements of channel prediction. Similarly, this necessitates different decision-making processes. Thus, DTC should be designed and activated in a task-oriented manner to enhance accuracy and efficiency.

 \subsection{DTC Modules and Enabling Techniques}
As depicted in Fig. \ref{fig_1}, DTC is driven by five key modules for multi-modal data collection, data processing and modeling, channel prediction, action decision, and real-time interaction, which will be detailed in the following

 \subsubsection{Multi-Modal Data Collection}
As commented in \cite{wang2023towards}, two types of information are mainly required: environment information and channel information. The physical environment encompasses extensive and differentiated attributes that cannot be fully captured by a single sensing modality. Different sensing approaches complement each other for comprehensive feature collections, as they sense the environment via different mechanisation and perspectives\cite{alkhateeb2023real}. Thus multi-modal sensing, comprising technologies like cameras and radar for measuring object locations, shape, and materials, satellite images for terrain and hydrology analysis, and sensors for climate recognition, is the crucial component in DTC.
\par In addition to environment data, DTC also requires multi-dimensional channel characteristics, including large-scale features like path loss and small-scale features like multi-path angles. During the DTC construction period, coarse channel parameters can be acquired beforehand through channel measurements or RT, and subsequently stored in a database. In the DTC implementation phase, real-time channel state information (CSI) can be obtained with various wireless devices and distributed infrastructures.

 \subsubsection{Data Processing and Modeling}
The collected heterogeneous raw data cannot be directly used for channel prediction since it typically suffers from undesired noises, redundant information, and a large data volume. Machine learning (ML) and data mining techniques can be employed to enable efficient denoising, predominant feature extraction, and dimensionality reduction. The processed online data can be categorized into instantaneous information and stationary information. The former describes the rapidly changing characteristics, such as relative locations of environment scatterers and real-time CSI. The stationary information describes the large-scale environment and channel features. Based on the macro-level data and database, different channel models can be established to characterize both the statistical and deterministic radio features, such as reflection, transmission, and statistical characteristics. Similarly, one can efficiently model the stationary environmental features, such as scatterer geometry, and implement 3D reconstructions combined with an environmental database.

 \subsubsection{Channel Prediction}

Based on the obtained multi-modal data, this module aims to construct powerful DMDD neural networks for realistic channel prediction. Since it is a task-oriented and scenario-defined problem, not all data are required. For instance, large-scale parameters (LSPs) and line-of-sight (LOS) blockage play predominant roles in evaluating channel quality, while small-scale parameters (SSPs) are less relevant. Therefore, it is necessary to collect service requests from the physical world and implement task-oriented feature selections aimed at enhancing information density. Then, multi-modal data fusion is required to help the constructed network in learning and inference. It can be facilitated by the deep learning (DL) technique, which enables hidden information mining and multi-level embedding representations of data. All the embedding representations will be concatenated and taken as input for the channel prediction network.

\par In this module, one should carefully balance the performance and complexity of the prediction network. Overfitting, robustness, and dataset requirements are common challenges in this field. Deep neural network (DNN) has shown promise as a strategy for predicting channel characteristics\cite{jiang2020deep}.  Typically, various heterogeneous DNNs should be trained for different communication tasks and scenarios to improve the generalization. From the model-driven perspective, a radio environment knowledge pool (REKP), comprising the knowledge of propagation laws abstracted from extensive channel measurements, should be established in advance \cite{wang2023towards}. It helps DNN comprehend and learn the underlying mapping rules of channel behaviors, enabling appreciable performance improvement and complexity reduction. According to the specific task and obtained stationary information, the associated knowledge will be selected and incorporated into the construction of DNN. For instance, the spatial consistency law can narrow down the effective search space to derive the propagation angle. Moreover, the channel statistical distribution can be taken into account for time-varying channel prediction, enabling higher accuracy compared to the data-driven AI \cite{zhang2023ai}.

 \subsubsection{Action Decision}
\par Using the predicted channel data, this module calculates optimized decisions for the users being served in the physical world. The predominant requirement is efficiency since the suggested actions should be timely transmitted to the wireless devices. It can be supported by the advancements of ML applications in the communication community \cite{nie2022predictive}. The key lies in training a set of heterogeneous ML networks capable of efficiently mapping the predicted channels to optimal decisions for different tasks. Similar to channel prediction, radio knowledge can be leveraged to facilitate optimal decision-making. Additionally, the suggested action should be translated into simple control commands to meet the real-time requirements. Utilizing the query-response structure can help simplify the transmission complexity from the digital world to the physical world.

 \subsubsection{Real-time interaction}
Once the decision-making is completed, both the suggested decisions and predicted channel characteristics will be transmitted to the physical network. This information facilitates a “coarse-refine” decision optimization for the served devices. Specifically, the channel prediction results provide a coarse awareness of the experienced channel for devices, while the suggested decisions help narrow down the decision-making options. Subsequently, the devices can further refine their actions by leveraging their own decision-making capability and collected information. 
\par For the feedback link, physical devices will send the real-experienced channel state data and derived decisions to the digital world. All the components of DTC such as the employed techniques, algorithm parameters, architecture, and assisted laws, can then be updated to improve the approximation of the physical world. \textcolor{black}{Note that physical devices might produce “dirty” data with incorrect and missing information. Thus pre-processing in the digital world should be conducted for abnormal data detection and cleaning \cite{rodriguez2023updating}. Moreover, the DTC model can be self-updated through federated learning (FL) by aggregating information from multiple nodes.}
\begin{figure}[t] 

 	\centering  
	\vspace{-0.1cm} 
	\subfigtopskip=-5pt 
	\subfigbottomskip=0pt 
	\subfigcapskip=-6pt 
	\setlength{\abovecaptionskip}{0.00cm}   
	\subfigure[\hspace{-3.5mm}]{
		\includegraphics[height=2.4cm]{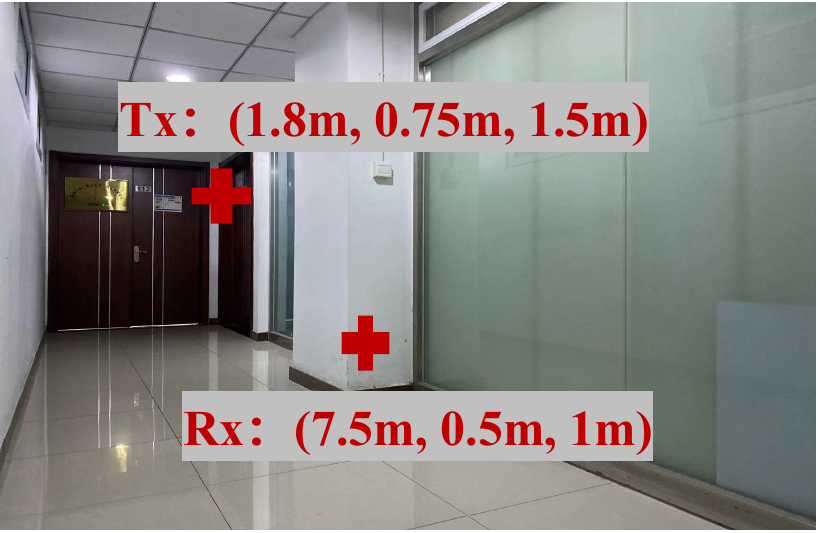}}\quad
  	\subfigure[\hspace{-3.5mm}]{
		\includegraphics[height=2.4 cm]{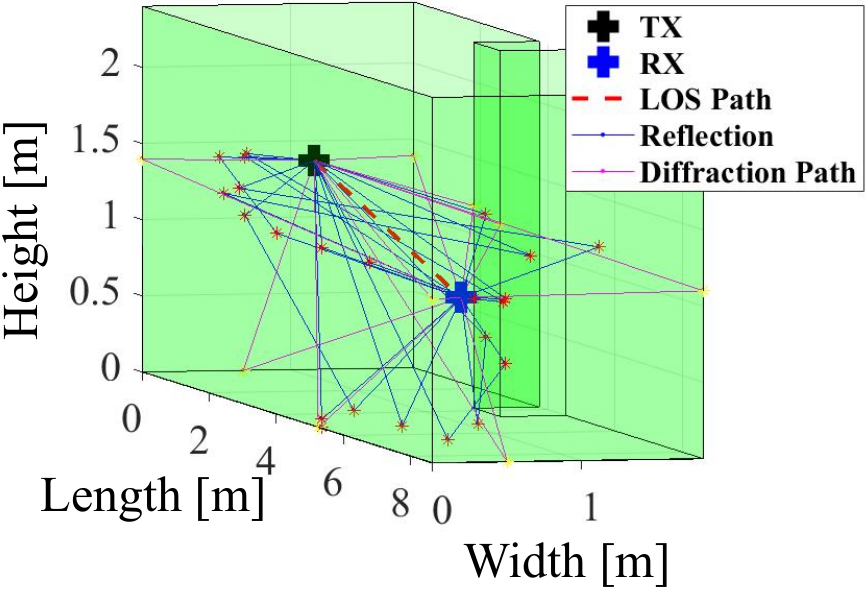}}
  	\subfigure[\hspace{-3.5mm}]{
		\includegraphics[width=4.1cm]{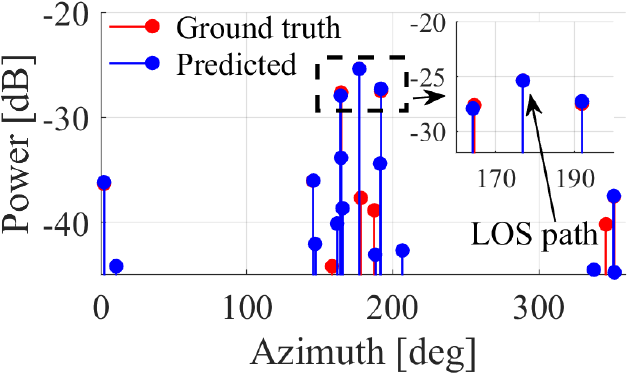}}
	\subfigure[\hspace{-3.5mm}]{
		\includegraphics[width=4.1cm]{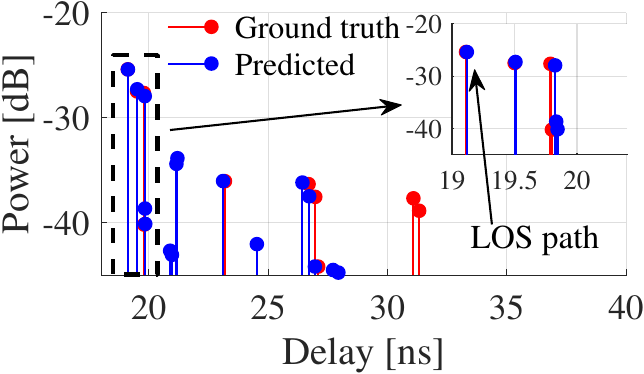}}
	\caption{\textcolor{black}{An example of channel prediction with the proposed platform.}}
	\label{Platform}
\end{figure}
\begin{figure*}[!t]
	\setlength{\abovecaptionskip}{-0.05cm}   
\centering
\includegraphics[width=5.2in]{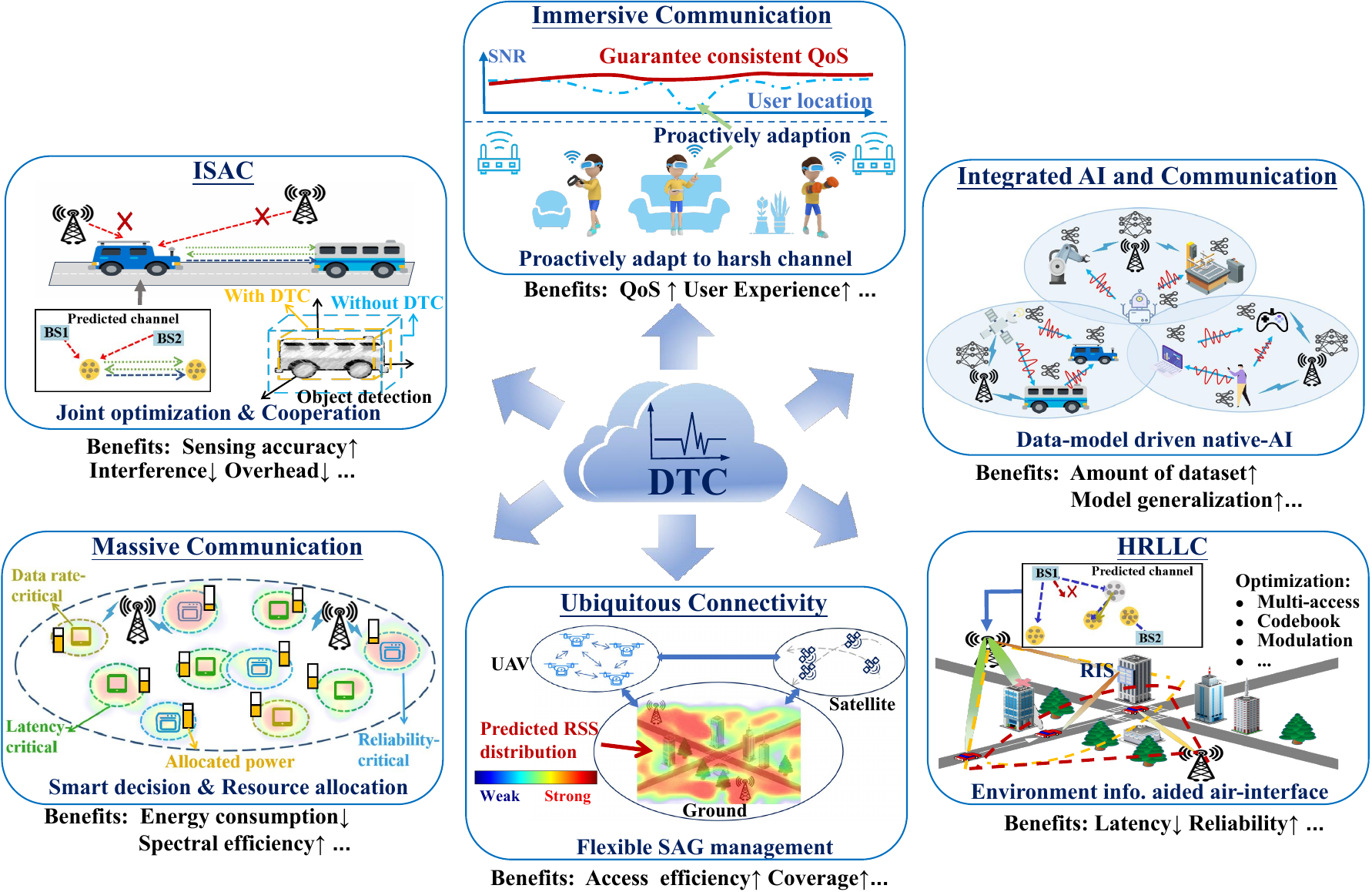}
\caption{The DTC enabled application prospects in 6G.}

\label{application}
\end{figure*}
 \subsection{Experimental Validation }
 \begin{figure}[!t]
 	\setlength{\abovecaptionskip}{-0.05cm}   
\centering
\includegraphics[width=3.4in]{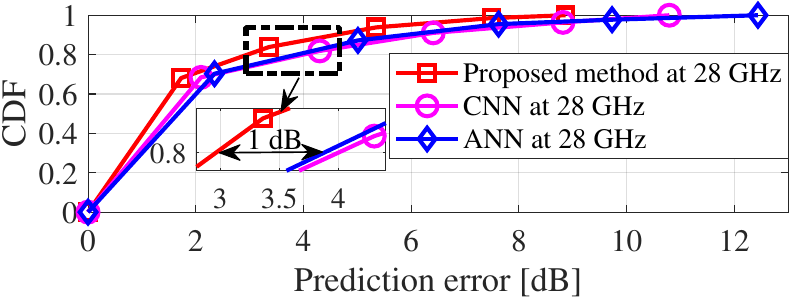}
\caption{Path loss prediction errors with different methods\cite{sun2022environment1}.}
\label{fig_4}
\end{figure}

In this part, we build a sensing-enhanced radio environment prediction system to evaluate the feasibility of real-time channel prediction\cite{miao2023demo}. \textcolor{black}{ Specifically, the environment sensing module collects real-time environmental information through sensing devices, generating point cloud data. Fig. \ref{Platform} (a) illustrates the sensing scenario, assuming a virtual Tx-Rx pair located at coordinates $(1.8m, 0.75m, 1.5m)$ and $(7.5m, 0.5m, 1m)$, respectively. The channel predictions are completed within seconds, with the results shown in Fig. \ref{Platform} (b)-(d). For comparison, the commercial Wireless InSite software is used to compute accurate channel parameters as ground truth values. As shown in Fig.\ref{Platform} (b), the predicted paths with the top three powers match well with the ground truth, with maximum power and delay deviations of $0.3$ dB and $0.04$ ns, respectively. Although there are some noticeable deviations, the paths with the top three powers contain $84.4\%$ of the total channel power, thereby dominating the channel characteristics and enabling effective communication decisions. Similarly, in the power-angle of arrival (AoA) domain, the predicted paths with the top three powers also match well, with the ground truth with maximum power and AoA deviations of $0.3$ dB and $0.3^{\circ}$, respectively, as shown in Fig.\ref{Platform} (d), demonstrating the accuracy of predicted channel characteristics.  }

\textcolor{black}{ The channel twin realized by the presented platform is between the $L_3$ and $L_4$. For an $L_5$ autonomous twin, the REKP and decision-making module \cite{wang2023towards} should be integrated into our platform to provide more reliable channel prediction and decision assistance for the transceiver. Additionally, a feedback link will be established to enable self-updates for the designed twin model. To further validate the feasibility of DTC, a more realistic prototype should be developed, and more practical communication tasks should be examined \cite{zeng2024tutorial}.}

\section{Applications}
DTC can facilitate the 6G system paradigm shift from non-scenario-specific design to scenario-specific design. Fig. \ref{application} showcases the typical 6G usage scenarios supported by DTC.
\par\subsubsection{Immersive Communication}
Compared to communication with two-dimensional (2D) images on a flat screen, immersive communication is expected to provide 3D stereoscopic images with audio-haptic information, which imposes demanding requirements on data transmission and reliability. \textcolor{black}{In these scenarios, DTC can be employed to predict link conditions in advance, such as path loss and communication disruptions. Based on the predicted link conditions and decision assistance from DTC, the immersive device can proactively adapt to various harsh channel conditions and ensure satisfactory data rates, network capacity, and user experiences. For this target, we designed an environment features-based path loss prediction method for DTC \cite{sun2022environment1}. Compared to the image-based conventional convolutional neural network (CNN) and raw data-based artificial neural network (ANN) methods, the proposed method is implemented with low-dimensional environment features such as blockage and scatter volume. This prediction method not only reduces the requirements for data and sensing hardware but also requires less computing time \cite{sun2022environment1}. Furthermore, as depicted in Fig. \ref{fig_4}, our method demonstrates more than $1$ dB performance improvement at an $80\%$ Cumulative Distribution Function (CDF) point compared to conventional methods. The prediction error is defined as the Root Mean Squared Error (RMSE) between the predicted results and the ground truth, both expressed in dB form.}

\par\subsubsection{Integrated Sensing and Communications (ISAC)}
As one of the typical usage scenarios in 6G, ISAC presents an attractive capability for emerging applications such as intelligent transportation, smart manufacturing, and multi-function radio frequency systems. One crucial concern in ISAC is the performance balance between sensing and communication. \textcolor{black}{With accurate channel prediction, adaptive pilot optimization can be conducted to perform low-overhead channel estimation and target detection, improving spectral efficiency and sensing accuracy.} Site-specific waveform optimization enabled by DTC promises significant improvements in both communication performance and sensing accuracy. Additionally, considering energy efficiency, DTC can assist in properly allocating power between communication and sensing modules. \textcolor{black}{Moreover, for network-level ISAC, the joint decision-making capability of DTC enables more optimized coordination and management, thereby enhancing multi-site interference mitigation and sensing performance.} Beyond these, more ISAC implementation tasks can be facilitated by DTC as well, such as beam optimization, joint operation, and others.
\par\subsubsection{Massive Communication}
As the transmission data rate, latency, and other communication requirements are expected to significantly increase in 6G, efficient optimization of network operations is crucial to support the massive number of wireless devices with diverse service requirements. Leveraging the environment information and decision-making assistance offered by DTC, network resources can be flexibly allocated for real-time communication services. Consider a scenario where terminals are randomly distributed across various cells. The terminals have diverse requirements, i.e., low-latency services, high-data-rate services, or highly reliable services. Resource blocks should be carefully allocated to different terminals to simultaneously maximize total throughput and respond to on-demand short-latency services in real time. \textcolor{black}{Since centralized DTC has a global view of the communication environment, the orientation and power of the base station (BS) array, resource block allocation, and the connections between terminals and different BSs can be jointly optimized for improving the spectral and energy efficiency.} Furthermore, by leveraging the stored historical network data, monitored current network status, and predicted future link conditions, DTC can support the proactive adaptation of terminals, enhancing communication reliability in dynamic physical environments. 
\par\subsubsection{Ubiquitous Connectivity}
As a complement to terrestrial systems, non-terrestrial networks play essential roles in the 6G ubiquitous connectivity scenario. In recent years, several mega-constellations, such as those by Telesat, OneWeb, and SpaceX, have been launched, with the aim of providing global coverage, especially for remote and disaster-affected areas. The advancement of unmanned aerial vehicle (UAV) technology facilitates flexible network redeployment, specifically for dynamic data traffic, thereby compensating for the limitations of fixed base stations. However, due to limited radio resources and extensive coverage, dynamic access management remains a critical issue in space-air-ground (SAG) communication. In this vision, according to the predicted channel characteristics and service requirements, DTC-based SAG systems can intelligently manage access conflict control and adjust the number of accessing terminals. \textcolor{black}{This approach could significantly enhance the access efficiency of SAG systems with dynamic terminal distribution.} Furthermore, as discussed before, the centralized DTC can provide joint decisions for diverse wireless nodes, thereby improving network performance.
\par\subsubsection{Hyper-Reliable and Low-latency Communication (HRLLC)} \textcolor{black}{With accurate channel prediction and intelligent decision-making, DTC plays significant roles in air-interface tasks such as multi-access, beam management, and handover. References \cite{zhang2023ai} and \cite{nie2022predictive} demonstrate the feasibility of predicting time-varying channel characteristics using low-dimensional environmental features and current channel status. This predictive capability allows the radio access network (RAN) to operate more intelligently and responsively, facilitating decisions like proactive handovers. Accurate channel information also enables scenario-specific air-interface strategies, such as adaptive modulation and multiple access designs. Moreover, by leveraging sensing and channel information from multiple nodes, DTC provides a global view of the communication environment. This capability empowers the RAN to make more sophisticated decisions. Illustratively, the sensed environment scatters, and predicted small-scale channel power parameters can aid in efficiently finding the optimal beam. Additionally, the predicted channel AoA contributes significantly to the multi-user interference mitigation. In essence, DTC supports a new paradigm in air-interface design by leveraging environmental information to enhance connection reliability, spectral efficiency, transmission rates, and more.}

\subsubsection{Integrated AI and Communication}
The native AI architecture represents a new paradigm in wireless networks that deeply integrates communication and AI techniques, providing differentiated AI services for various application scenarios. Accessing rich real-world datasets spanning network operations, channel characteristics, and physical environments is crucial for the native AI architecture. \textcolor{black}{The obtained AI model may suffer from an over-fitting problem if the training data is insufficient. However, it is laborious to perform extensive measurements and data collection in practice. The generative adversarial networks (GAN)-based channel prediction can generate massive data without the need for practical measurements. This technique can be exploited in the DTC for channel prediction/inference. \textcolor{black}{Since the obtained dataset has the same statistical properties, the over-fitting problem can be effectively mitigated.} Furthermore, since DTC is expected to serve various communication devices and collect multi-modal information across different environments, it can learn and predict channel characteristics covering various wireless scenarios, such as indoor, outdoor, and industrial Internet of Things. Therefore, one can generate various channel datasets for diverse wireless scenarios using DTC.}
\textcolor{black}{
\section{Open Issues}
Tough promising, DTC is still at the beginning of industrial application. Various open issues should be addressed moving forward, which can be categorized into three main areas.}
\par\textcolor{black}{ \textbf{Construction of DTC:} Accurate mapping from the physical world to the digital world is the foundation of DTC, which requires a significant amount of data during the DTC construction period. The main concern is how to cost-effectively obtain a massive amount of data for DTC model training. Additionally, efficiently building and storing a set of heterogeneous models for diverse communication tasks needs to be addressed. Radio knowledge is essential for the task-oriented DMDD framework. Further investigation is required on how to design a generalized REKP for diverse communication tasks. }

\par \textcolor{black}{\textbf{Utilization of DTC:}  Real-time interaction involves transmitting multi-modal information and correction commands between the physical and digital worlds, along with conveying suggested decisions and predicted channel information. This increases communication overhead. More studies are needed to optimize the interaction manner to improve the efficiency. Furthermore, substantial computational resources are required to process complex data and algorithms in real time, making the design of a network with adequate computing power one of the critical challenges. Additionally, the sharing of information during utilization may raise concerns about privacy and security \cite{zeng2024tutorial}.
}

  \textcolor{black}{\textbf{Self-updating mechanism:} DTC is not always highly accurate in approximating the physical world, making self-updating critical for maintaining accurate digital models. However, retraining the entire DTC model is impractical. \cite{zeng2024tutorial} pointed out that incremental learning algorithms can be employed to update the system model efficiently without the need for complete retraining. Moreover, efficient methods for abnormal data detection and FL-based digital model correction should be further investigated. More relevant studies and experimental validation are required.
}

\section{Conclusion}

This article proposes a real-time DTC framework for a 6G network and presents the associated key enablers. Firstly, we define five levels of channel twins and specify the key features. Subsequently, the requirements and architecture for DTC are discussed in detail. Additionally, we experimentally evaluate the feasibility of real-time environment sensing and wireless channel prediction, the results of which align well with the ground truth. Finally, we discuss the potential applications of DTC in 6G usage scenarios, as well as the open issues.


\section*{Acknowledgments}
This work is supported by the National Key R\&D Program of China (Grant No. 2023YFB2904803), the National Science Fund for Distinguished Young Scholars (No.61925102), the National Natural Science Foundation of China (No.62341128, No.92167202), BUPT-CMCC Joint Innovation Center, and the Postdoctoral Fellowship Program of China Postdoctoral Science Foundation (CPSF) under Grant No. GZB20230087.

 \bibliographystyle{IEEEtran}
 \bibliography{IEEEabrv,ref}

\vspace{-35pt}

\begin{IEEEbiographynophoto}{HENG WANG  (hengwang@bupt.edu.cn)}
 is currently a postdoctoral research fellow with Beijing University of Posts and Telecommunications. 
\end{IEEEbiographynophoto}
\vspace{-35pt}
\begin{IEEEbiographynophoto}{JIANHUA ZHANG  (jhzhang@bupt.edu.cn)}
 is currently a professor with Beijing University of Posts and Telecommunications. 
\end{IEEEbiographynophoto}

\vspace{-35pt}

\begin{IEEEbiographynophoto}{GAOFENG NIE  (niegaofeng@bupt.edu.cn)}
 is currently a associate professor with Beijing University of Posts and Telecommunications. 
\end{IEEEbiographynophoto}

\vspace{-35pt}

\begin{IEEEbiographynophoto}{LI YU  (li.yu@bupt.edu.cn)}
 is currently a postdoctoral research fellow with Beijing University of Posts and Telecommunications.  
\end{IEEEbiographynophoto}

\vspace{-35pt}

\begin{IEEEbiographynophoto}{ZHIQIANG YUAN   (yuanzhiqiang@bupt.edu.cn)}
 is currently pursuing the Ph.D. degree with Beijing University of Posts and Telecommunications.
\end{IEEEbiographynophoto}

\vspace{-35pt}

 \begin{IEEEbiographynophoto}{TONGJIE LI   (2020110200@bupt.edu.cn)}
 is currently pursuing the Ph.D. degree with Beijing University of Posts and Telecommunications.
\end{IEEEbiographynophoto}
\vspace{-35pt}

 \begin{IEEEbiographynophoto}{JIALIN WANG   (wangjialinbupt@bupt.edu.cn)}
 is currently pursuing the Ph.D. degree with Beijing University of Posts and Telecommunications.
\end{IEEEbiographynophoto}

\vspace{-35pt}

\begin{IEEEbiographynophoto}{GUANGYI LIU  (liuguangyi@chinamobile.com)}
 is  currently Fellow and 6G lead specialist with China Mobile Research Institute.
\end{IEEEbiographynophoto}

\vfill

\end{document}